\documentclass[11pt]{article}
\usepackage[utf8]{inputenc}
\usepackage[utf8]{inputenc}
\usepackage{amsfonts, amsmath, amsthm, amssymb} 
\usepackage{graphicx} 
\usepackage{hyperref} 
\usepackage{multirow,tabu} 
\usepackage{array}
\newcolumntype{H}{>{\setbox0=\hbox\bgroup}c<{\egroup}@{}}
\usepackage{chngcntr}
\usepackage{authblk}
\usepackage{color}
\usepackage{float}
\usepackage[backend=biber ,style=numeric,maxbibnames=9,natbib ]{biblatex}
\title{Effects of winter climate on high speed passenger trains in Botnia-Atlantica region}
\author{Jianfeng Wang\thanks{Corresponding author : Jianfeng Wang\\
\hspace*{4.3 mm} Email: jianfeng.wang@umu.se}}
\author{Markus Granlöf}
\author{Jun Yu}
\affil{Department of Mathematics and Mathematical Statistics, Umeå University, SE 901 87 Umeå, Sweden}
\date{}

\begin{document}

\maketitle

\begin{abstract}
Harsh winter climate can cause various problems for both public and private sectors in Sweden, especially in the northern part for railway industry. To have a better understanding of winter climate impacts, this study investigates effects of the winter climate including atmospheric icing on the performance of high speed passenger trains in the Botnia-Atlantica region. The investigation is done with train operational data  together with simulated weather data from the Weather Research and Forecast model over January - February 2017.

Two different measurements of the train performance are analysed. One is cumulative delay which measures the increment in delay in terms of running time  within two consecutive measuring spots, the other is current delay which is the delay in terms of arrival time at each measuring spot compared to the schedule. Cumulative delay is investigated through a Cox model and the current delay is studied using a Markov chain model.

The results show that the weather factors have impacts on the train performance. Therein temperature and humidity have significant impacts on both the occurrence of cumulative delay and the transition probabilities between (current) delayed and non-delayed states.
\end{abstract}


\section{Introduction}

Cold, thick snow and atmospheric icing are three well known winter phenomenon in Nordic countries. Such climate puts a strain on infrastructure, companies and people, which affects the normal operations of the whole society. It is therefore central to gather knowledge and methods to counter problems caused by the extreme winter climate. In this study, the effect of harsh winter climate on railway system is investigated.

The railway system is an important part of the infrastructure and a large number of people as well as companies depend on it for transportation on sorts of purposes every day. Thus, punctuality becomes one key criterion for the railway industry  in order to minimise the society costs and increase its reliability of the railway system. To achieve this goal, an important task is therefore to investigate and figure out how train delays are affected by winter climate factors.

The performance of the train operations is commonly measured by cumulative delay and current delay. Cumulative delay measures the increment in delay within two consecutive measuring spots in terms of running time, and current delay is the delay in terms of arrival time at each measuring spot compared to the schedule.

Earlier studies about train performance and the effect of weather have been conducted. \citet{Xia2013} fitted a linear model and showed that weather factors like snow, temperature, precipitation and wind had significant effects on the punctuality of trains in Netherlands. In a more recent study by \citet{Wang2019}, a machine learning approach was used to create a predictive model to predict the current delay at each station for a train line in China with help of weather observations. \citet{Ottosson} used count data models such as negative binomial regression and zero-inflated model and showed that weather variables, such as snow depth, temperature and wind direction, had significant effects on the train performance.

 The cumulative delay information of a train trip collected from the measuring spots in time could also be seen as time-to-event (i.e. cumulative delay) data, and the transitions between the (current) delayed and non-delayed states in a train trip can be treated as a Markov chain. To the best of our knowledge, very few applications of the survival analysis to the railway field can be found in literature, and they do not concern the delay issues \citep{Jardine,Annelise,Andersson}; there are also a few applications of Markov chain to the train delay investigation, but none of them considered the weather impact on the transition probability \citep{Sahin,Kecman,Gaurav}.

Therefore, novelty of this study is that survival analysis is used to investigate how the winter climate affects the occurrence of cumulative delays, and Markov chain model is applied to study the effect of winter climate on the transitions between delayed and non-delayed states.

The weather data is simulated from the Weather Research and Forecasting (WRF) model instead of using real meteorological observations, since the distance between the nearest meteorological station and measuring spot along the train line ranges from 17 to 24 km \citep{Ottosson}. Thus, using the meteorological data is not an ideal choice in the analysis. However, WRF with high spatial resolution is a good alternative under this situation, which is one of the most commonly used numerical weather prediction models for both atmospheric research and operational forecasting needs. Its reliable performance has been assessed in a number of studies \citep{wangbayesian, wangdownscal,Mohan,Cassano}.

The paper is organized as follows. In Section 2, we introduce the statistical models in details. Data processing and  model diagnostic methods are described in Section 3. Section 4 is reserved for results. Section 5 is devoted to the conclusion and discussion.

\section{Statistical modelling}
\subsection{Cox proportional hazard model with time dependent covariates for recurrent event}
\citet{Andersen1982} proposed a regression model used to analyse the relationship between survival times (or hazard rates) of subjects with recurrent event and covariates, which is probably the most often applied model for recurrent event survival analysis. It is a simple extension of the original Cox model \citep{Cox}. Formally, the extended Cox model with time dependent covariates for recurrent event is an expression of the hazard function $h(t)$, which gives the risk of an event at time \emph{t}, and covariates

\begin{equation}
\label{coxmodel}
   h_{ij}(t)  = h_0(t)\exp{(\boldsymbol{\beta}^T \mathbf{x}_{ij}(t))},
\end{equation}
where
\begin{itemize}
\item $h_{ij}(t)$ represents the hazard function for the $j$th event of
the $i$th subject at time $t$.
\item $h_0(t)$ is the baseline hazard which is the hazard rate when all the covariates are equal to zero.
\item $\mathbf{x}_{ij}(t)$ represents a covariate vector for the $i$th subject and the $j$th event at time $t$.
\item $\boldsymbol{\beta}$ is an unknown  coefficient vector to be estimated.
\end{itemize}

The coefficients can be estimated with partial likelihood by taking into account the conditional probabilities for the events that occur for subjects, which is given by
\begin{equation}
    \label{partialLikli}
    L(\boldsymbol{\beta})  = \prod_{i = 1}^n \prod_{j = 1}^{k_i} \left(\frac{\exp{( \boldsymbol{\beta}^T \mathbf{x}_{i}(t_{ij})))}}{\sum_{l \in R(t_{ij})} \exp{( \boldsymbol{\beta}^T \mathbf{x}_{l}(t_{ij})})}\right)^{\delta_{ij}},
\end{equation}
where $j$ is the event index with $k_i$ being the subject-specific maximum number of events, $\mathbf{x}_{i}(t_{ij})$ denotes the covariate vector for the $i$th subject at the $j$th event time $t_{ij},$ $\delta_{ij}$ is an event indicator which equals $1$ for the $j$th event of the $i$th subject and $0$ for censoring, $R(t) = \{l, l=1,\cdots,n: \exists  j\in\{1,\cdots,k_l\} \text{ with } t_{lj}\ge t\}$ is a group of subjects that are at risk for an event at time $t$.

The proportional hazard assumption is a vital assumption to the use and interpretation of a Cox model, and it states that the hazard ratio of two subjects, $i$ and $l$, at time $t$
\begin{equation}
\label{hr}
    HR(t)=\frac{h_{i}(t)}{h_{l}(t)}=\exp{\left(\boldsymbol{\beta}^T\left(\mathbf{x}_{i}(t)-\mathbf{x}_{l}(t)\right)\right)}
\end{equation}
is constant. An obvious case that (\ref{hr}) is constant when the covariate vectors $\mathbf{x}_i(t)$ and $\mathbf{x}_l(t)$ are independent of $t$. In general, the proportionality assumption may be invalid for time dependent covariates. A likelihood ratio test or Schoenfeld residuals can be used to examine whether a covariate satisfies the assumption or not \citep{Abeysekera}. If not, a heaviside function $g(t)$ can be chosen to fix the proportionality violation \citep{Miftahuddin}. The purpose of a heaviside function is to partition the observational time into intervals and make the proportionality assumption be valid within each interval. Then (\ref{coxmodel}) could be rewritten element-wise as
\begin{equation}
\label{heaviside}
   h_{ij}(t)  = h_0(t)\exp{\left(\sum_{m=1}^p\beta_m x_{ijm}+\sum_{m=1}^p\theta_m x_{ijm}g_{m}(t)\right)},
\end{equation}
where $x_{ijm}$ is the $m$th covariate of the $i$th subject and the $j$th event, $\theta_m$ is a new introduced coefficient for the $m$th covariate in an interval defined by $g_m(t)$, and $g_m(t)$ is a function of time for the $m$th covariate. For the case, $g_m(t)=0$ for all $m$ implies no time dependent covariates, then (\ref{heaviside}) is reduced to a simpler version of (\ref{coxmodel}) with only time independent covariates
\begin{equation}
\label{timeinde}
   h_{ij}(t)  = h_0(t)\exp{(\boldsymbol\beta^T\mathbf{x}_{ij})}.
\end{equation}

In this study, a simple step function is chosen for covariates that do not satisfy the proportionality assumption
\begin{equation}
\label{heav}
 g_m(t) = \Bigg\{
  \begin{array}{l c }
    0, \  t \ \leq t_0,\\
     1, \  t \ >t_0,
 \end{array}
\end{equation}
where $t_0$ is chosen through Schoenfeld residuals to make the proportionality assumption hold within each interval.

\subsection{Markov chain model}
A multi-state model describes how an subject moves among a number of states in continuous time \citep{Jackson}. The movement from state $r$ to $s$ at the time $t$ is governed by transition intensity
\begin{equation}
\label{inte}
    q_{rs}(t) = \lim_{\Delta t \to 0} P( S(t+\Delta t) = s |S(t)=r)/\Delta t,
\end{equation}
which may depend also potentially on a time dependent explanatory vector $\mathbf{x}(t)$, i.e. $q_{rs}(t)\to q_{rs}(t,\mathbf{x}(t))$.

The $q_{rs}$ of a $q$ states process forms a $q\times q$ transition intensity matrix $Q$, whose rows sum to zero, so that the diagonal entries are defined by $q_{rr} = -\sum_{s\neq r}q_{rs}$. An example of transition intensity matrix $Q$ with two states can be seen below

\begin{equation}
Q = \begin{bmatrix}
    q_{11} & q_{12} \ \\
    q_{21} \ & q_{22} \end{bmatrix},
\end{equation}
where $q_{11}=-q_{12}$ and $q_{22}=-q_{21}$.

A multi-state model generally relies on the Markov assumption that next transition only depends on the current state, i.e. $q_{rs}(t, \mathbf{x}(t),\mathcal{H}_t)$ is independent of $\mathcal{H}_t$ which is the observation history of the process up to the time preceding $t$.

At the early investigate of the study, we limit our focus on a homogeneous process in time meaning that the transition intensity $Q$ is constant and the transition probability to move from a state to another depends solely on the time difference between two time points, i.e.

\begin{equation}
    \label{mark2}
    P( S(u+t) = s |S(u) = r) = P(S(t) = s | S(0) = r).
\end{equation}

Corresponding to the transition intensity matrix $Q$, the entry in a transition probability matrix $P(u,u+t)$ is $p_{rs}(u,u+t)$ representing the probability of being in state $s$ at a time $u+t$ given the state at time $u$ is $r$. The relationship between transition intensity matrix and transition probability matrix is defined through the Kolmogorov differential equations \citep{Cox1977}. Specially, when a process is homogeneous, the transition probability matrix can be calculated by taking the matrix exponential of the transition intensity
matrix
\begin{equation}
\label{pro}
    P(u,u+t)=P(t) = \text{Exp}(tQ).
\end{equation}

In a Markov chain model, to take account of the effects of explanatory variables, a  proportional hazard like model was proposed by \citet{Marshall1995}
\begin{equation}
\label{explo}
    q_{rs}(t,\mathbf{x}(t)) = q_{rs}^{(0)}(t) \exp{(\boldsymbol{\beta}_{rs}^T \mathbf{x}(t))},
\end{equation}
where $q_{rs}^{(0)}(t)$ is baseline transition intensity when all covariates are zero.

The coefficient vectors $\boldsymbol{\beta}_{rs}$ as well as the transition intensity matrix $Q$ and the transition probability matrix $P(t)$ can be estimated through maximising the likelihood

\begin{equation}
    \label{logts}
    L(Q) = \prod_{i,j} p_{S(t_{i,j})S(t_{i,j+1})}(t_{i,j + 1}- t_{i,j}),
\end{equation}
where the transition probability is evaluated at time $t_{i,j + 1}- t_{i,j}$ and $S(t_{i,j})$ represents the $j$th observed state of the $i$th subject at time $t_{i,j}$.
\section{Method}
\subsection{Train data}
The investigation focuses on the high speed passenger train, which is a type of trains with top speed of between 200 to 250 km/h. This type of train often travels longer distances and is therefore more prone to experience disturbances caused by climate factors. A train line comprises of a number of measuring spots where the operational times are recorded such as departure and arrival times.

In the study, the train line between Umeå and Stockholm is selected, which includes 116 measuring spots in total. The total length of the train line is 711 km and the planned drive time for a high speed passenger train is approximately 6.5 hours. The lengths of any two consecutive measuring spots vary from 0.3 km to 15 km. The train operation data in the year 2017 is provided by the Swedish Transport Administration. The key variables are listed in Table \ref{table:1}.
\begin{table}[H]
\centering
\caption{List of variables in the train operation data}
\begin{tabular}{ | m{11em} | m{9cm}|  } \hline
\textbf{Variables}  & \textbf{Description} \\ \hline
Train Number & An identification number for train used in the trip  \\ \hline
Arrival location & Name of arrival measuring spot  \\ \hline
Departure location & Name of departure measuring spot \\ \hline
Departure date &  The departure date (yyyy-mm-dd) for a train at a location.  \\ \hline
Arrival date &  The arrival date (yyyy-mm-dd) for a train at a location. \\ \hline
Train type &  Type of train, for example: high speed, commute train and regional \\ \hline
Section Length  & Length (km) between two consecutive measuring spots \\ \hline
Planned departure time  & The planned departure time (hh:mm) at a measuring spot \\ \hline
Planned arrival time  & The planned arrival time (hh:mm) at a measuring spot \\ \hline
Actual departure time  & The Actual departure time (hh:mm) at a measuring spot\\ \hline
Actual arrival time  & The Actual arrival time (hh:mm) at a measuring spot \\ \hline
\end{tabular}
\label{table:1}
\end{table}

\subsection{Weather data}
\begin{figure}[H]
    \centering
    \includegraphics[scale=0.7]{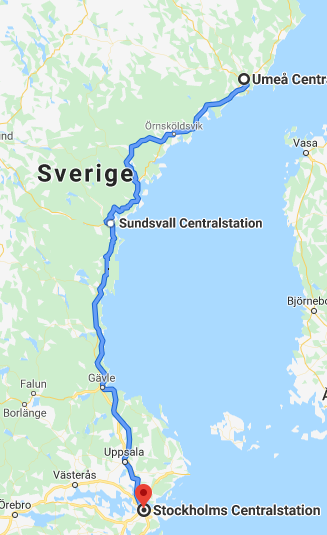}
    \caption{Train line in the region with simulated WRF data}
    \label{pic:weather}
\end{figure}

 A WRF model is a numerical weather prediction system that is used for research and operational purposes. The model can be set up with different configurations and over different regions. Actual atmospheric conditions and idealised conditions can be used in the model. The WRF model simulates desired weather variables estimations over grids. Higher spatial resolution gives smaller square of each grid. Temporal resolution decides the time interval between each simulation. The extreme winter weather data from January to February in the year 2017 is of special interest in the study. The spatial resolution is $3\times3$ km and the temporal resolution is 1 hour. The simulation region as well as the train line in blue of interest are shown in \textit{Figure \ref{pic:weather}}.

The weather variables of interest are shown in \textit{Table \ref{table:2}}. These variables are chosen because they are believed to have impacts on the train operation in winter and we want to investigate how these variables affect the train operation.

\begin{table}[H]
\centering
\caption{The weather variables of interest}
\begin{tabular}{ | m{8em} | m{10cm}|  }
\hline
\textbf{Variables}  & \textbf{Description} \\
\hline
Temperature  & The temperature ($^\circ$C) at 2 meters above the ground \\  \hline
Humidity  & Relative Humidity (\%) at 2-meters   \\  \hline
Snow depth & The snow depth in meters (m)  \\  \hline
Atmospheric icing  &  Accumulated snow and ice in millimeter (mm) \\  \hline

\hline
\end{tabular}

\label{table:2}
\end{table}
 Every measuring spot on the train line is matched with the closest grid point by date and time. The measuring time in train operation data has to be rounded to the closest hour.

The mean of the weather variables within any two consecutive spots is calculated and used in the analysis. Since a large number of the atmospheric icing values is zero along the train line, a categorical variable is used instead of the continuous atmospheric icing variable, i.e. 0 if atmospheric icing is zero, 1 otherwise.

\subsection{Missing values in the train operation data}
A section between two consecutive measuring spots for a train trip often has missing departure/arrival times that can be classified into three different classes which are defined in \textit{Table \ref{table:miss}}.

\begin{table}[H]
\centering
\caption{Classes of missing times}
\begin{tabular}{ | m{8em} | m{5cm}| m{5cm}|   }
\hline
\textbf{Class}  & \textbf{Departure time missing}& \textbf{Arrival time missing} \\
\hline
1& True & False\\  \hline
2& False & True \\  \hline
3& True & True \\  \hline
\end{tabular}

\label{table:miss}
\end{table}

 A common method to impute missing values in such longitudinal data is called last observation carried forward (LOCF), i.e. the latest recorded value is used to impute the missing value.  The advantages of using LOCF are that the number of observations removed from the study decreases and make it possible to study all subjects over the whole time period. A disadvantage with the method is the introduction of bias of the estimates if the values changes considerably large with time or the time period between the most recent value and the missing value is long.
Because the intervals with missing values are short in the dataset which decreases the risk of bias, thus it is reasonable to apply this approach. Based on the LOCF, the imputation procedure is explained further below.

\begin{enumerate}
  \item Start from the beginning of the trip and save the latest arrival and departure time until a missing time is occurring.

  \item If the missing time is arrival time then (a); if departure time is missing then (b):
  \begin{enumerate}
     \item Replace the missing arrival time with the latest departure time + the planned driving time for the previous section
     \item Replace the missing departure time with the latest arrival time  + the planned dwell time.
     \end{enumerate}

    \item Save the imputed time as the new latest time.

    \item If the section is not the last section of the trip, go back to step 1.

\end{enumerate}

\subsection{Proportionality test}
It is necessary to verify whether the proportionality assumption holds for a fitted Cox model. A Schoenfeld residuals based test can be used to test the assumption. Schoenfeld residuals has the form \textit{Observed - Expected}. The Schoenfeld residual $r$ for the $j$th event is defined as
\begin{equation}
    \label{Schoenfeld}
    r_j(\hat{\boldsymbol{\beta}})=\mathbf{x}_i(t_{ij})-E(\mathbf{x}(t_{ij})|R(t_{ij}))=
    \mathbf{x}_i(t_{ij})-\sum_{l\in R(t_{ij})}\mathbf{x}_{l}(t_{ij})w_l,
\end{equation}
where $w_l=\frac{\exp{\left( \hat{\boldsymbol{\beta}}^T \mathbf{x}_{l}(t_{ij}))\right)}}{\sum_{l' \in R(t_{ij})} \exp{\left( \hat{\boldsymbol{\beta}}^T \mathbf{x}_{l'}(t_{ij})\right)}}$ is a weight of $\mathbf{x}_{l}(t_{ij})$ over those observations still at risk at time $t_{ij}$. If the proportional hazard assumption holds, $E(r_j(\hat{\boldsymbol{\beta}}))\approx 0$ \citep{Xue,Abeysekera}. \citet{harrell1979phglm} proposed a test based on the Schoenfeld residual. It is a test of correlation between the Schoenfeld residuals and time, for example, a correlation of zero indicates that the model met the proportional hazard assumption.
\subsection{Analysis tools}
R is the programming language that used throughout the study. Therein, the package \textit{dplyr} is used for data processing, the package \textit{survival} and the package \textit{msm} are used for the Cox model and Markov chain model, respectively.

\section{Results}
\subsection{Cox model}
By using the proportionality test, it shows that the temperature variable does not satisfy the proportionality assumption. After inspecting the Schoenfeld residuals plotted against travel distance (not shown), the hazard ratios are different between 0 - 150km and 150 - the end. Therefore, a heaviside function
\begin{equation}
\label{heavyy}
 g(t) = \Bigg\{
  \begin{array}{l c }
     0, \   t \ \leq 150,\\
     1, \   t \ >150,
 \end{array}
\end{equation}
is chosen for temperature.

The results from the fitted extended Cox proportional hazard model with (\ref{heavyy}) can be found in \textit{Table \ref{table:solocox}}. Two predictors have significant effects on the occurrence of the cumulative delay which are temperature and humidity. To be specific, as temperature increases 1 $^\circ$C, the hazard rates decrease 17\% for the first 150 km and 6\% for the remaining trip, respectively.  As humidity increase 1\%, the hazard rate increases 2.6\%.

\begin{table}[H]
\caption{Estimates from the fitted extended Cox model }
\centering
\begin{tabular}{ | m{6em} | m{2cm}|  m{2cm}|  m{2cm}|  m{2cm}|  m{2cm}| }
\hline
\textbf{Predictor}  & \textbf{Coefficient} & \textbf{Hazard ratio} & \textbf{Robust standard error} & \textbf{z-value} & \textbf{p($>$z)}  \\ \hline
Temperature (0 - 150 km) & -0.19& 0.83& 0.055  & -3.39 &0.00070*     \\\hline
Temperature (150 km - the end)& -0.062&  0.94& 0.025 &-2.46952& 0.014*     \\\hline
Humidity & 0.025&  1.026 & 0.0081 & 3.14 &0.0017*    \\\hline
Snow depth & 0.0073 & 1.0074 &  0.015  &0.48 &0.63     \\\hline
Categorical icing & -0.16& 0.85& 0.15  & -1.061 &0.29    \\\hline

\end{tabular}

\label{table:solocox}
\end{table}

The proportionality test for the model with the heaviside function for each predictor can be seen in \textit{Table \ref{proportionality}}. It shows the proportionality assumption holds for extended Cox model with the heaviside function.

\begin{table}[ht]
\centering
\caption{P-values from proportionality test}
\begin{tabular}{ | m{10em} | m{5cm}| }
\hline
\textbf{Predictor}  & \textbf{P-value}  \\ \hline
Temperature & 0.45    \\\hline
Humidity  &0.54 \\  \hline
Snow depth &0.51  \\  \hline
Categorical icing  &0.48 \\  \hline
\textbf{Global}  &0.78   \\ \hline
\end{tabular}

\label{proportionality}
\end{table}

\subsection{Markov chain model}
\textit{Table \ref{table:markpre}} shows temperature and atmospheric icing have significant impacts on the transition rate from non-delayed to delayed states in the model. The hazard ratios indicate that as the temperature increases 1 $^\circ$C, the transition rate from non-delay to delayed states decreases 3\%, and the transition rate from non-delay to delayed states increases 46\% with the occurrence of atmospheric icing.

\begin{table}[H]
\centering
\caption{Hazard ratios from non-delayed to delayed states}
\begin{tabular}{ | m{6em} | m{2cm}|  m{2cm}|m{2cm}|  }
\hline
\textbf{Predictor}  & \textbf{Hazard Ratio} & \textbf{CI: Lower}& \textbf{CI: Upper}   \\ \hline
Temperature & 0.97 &0.94&0.99\\\hline
Humidity &  0.99 &0.98 &1.002 \\\hline
Snow depth & 1 &0.97&1.024 \\\hline
Categorical icing & 1.460  & 1.163&  1.83 \\\hline

\end{tabular}

\label{table:markpre}
\end{table}

\textit{Table \ref{table:markov}} shows that the estimates of hazard ratios from delayed to non-delayed states. Temperature, humidity and snow depth are significant in the model. It indicates that as the temperature increases 1 $^\circ$C, the transition rate from delayed to non-delay states increases 3.3\%, as the humidity increases 1\%, the transition rate decreases 2\%, and as the snow depth increases 1 m, the transition rate decreases 5\%.
\begin{table}[H]
\centering
\caption{Hazard ratios from delayed to non-delayed states}
\begin{tabular}{ | m{6em} | m{2cm}|  m{2cm}| m{2cm}|}
\hline
\textbf{Predictor}  & \textbf{Hazard Ratio} & \textbf{CI: Lower}& \textbf{CI: Upper}    \\ \hline
Temperature & 1.033&1.0010&1.065  \\\hline
Humidity &  0.98 &0.96&0.99 \\\hline
Snow depth & 0.95& 0.92&0.99  \\\hline
Categorical icing & 1.19&0.945&1.50   \\\hline

\end{tabular}

\label{table:markov}
\end{table}

\section{Conclusion and discussion}
The purpose of this study was to investigate how the winter weather as well as the atmospheric icing affect the occurrence of cumulative delays and the transitions between delayed and non-delayed states. The cumulative delay was investigated with the Cox proportional hazard model with time dependent covariates for recurrence event, which showed that the temperature and humidity have significant effects on the occurrence of cumulative delays. Lower temperature and higher humidity increase the probability of the occurrence of cumulative delays.

The transitions between delayed and non-delayed states were investigated with the two-states Markov model. This showed that the temperature and the atmospheric icing have significant effects on the transition rate from non-delayed to delayed states. More specifically, lower temperatures and the presence of icing increase the transition rate from non-delayed to delayed states. In the other side, humidity, temperature and snow depth have significant effects on the transition rate from delayed to non-delayed states. To be specific, higher temperatures, lower snow depth and lower humidity increase the transition rate from delayed to non-delayed states.

In summary, both models show that the temperature and humidity have  impacts on the performance of a train in the winter climate, i.e. lower temperature and higher humidity are against the train to be punctual.

Much could be done in terms of statistical modelling in the further investigation. For instance, 1) the stratified Cox model can be used, which takes the order of event into account and could also avoid to violate the proportionality assumption \citep{Ozga}; 2) fitting an inhomogeneous Markov chain model to the train operation data is more reasonable, since it is strict to assume the transition rate does not change over time in reality; 3) a more than two states' Markov chain model can be used to acquire a deeper understanding about the climate effects. Besides, more train operation data as well as weather data could to be included in the model construction and verification procedure.

\section*{Acknowledgements}
We acknowledge EU Intereg Botnia-Atlantica Programme and Regional Council of V\"{a}sterbotten and Ostrobothnia for their support of this work through the noICE project. We would like to thank the Swedish Transport Administration for providing the train operation data. We would like to thank  the Atmospheric Science Group (ASG) at Lule\r{a} University of Technology for simulating and providing the WRF data.  

\printbibliography
\end{document}